\documentstyle[prl,twocolumn,aps,epsf]{revtex}
\begin{document}

\twocolumn[
\hsize\textwidth\columnwidth\hsize\csname@twocolumnfalse\endcsname

\title{Parallel magnetic field induced giant magnetoresistance in 
low density {\it quasi}-two dimensional layers}
\author{S. \ Das Sarma and E. H.\ Hwang}
\address{Department of Physics, University of Maryland, College Park,
Maryland  20742-4111 } 
\date{\today}
\maketitle

\begin{abstract}

We provide a possible theoretical explanation for the recently observed giant
positive magnetoresistance in high mobility low density {\it quasi}-two 
dimensional electron and hole systems.
Our explanation is based on the strong coupling of 
the parallel field to the {\it orbital} motion arising from the 
{\it finite} layer thickness and the large Fermi wavelength of the 
{\it quasi}-two dimensional system at low carrier densities.

\noindent
PACS Number : 73.50.Jt; 71.30.+h, 73.50.Bk; 73.40.Hm; 73.40.Qv

\end{abstract}
\vspace{0.5in}
]

\newpage

Recently an intriguing set of low temperature  transport
properties of low density high mobility two dimensional (2D) electron
(Si inversion layers) and hole (p-type GaAs heterostructures) systems, 
being collectively referred to as the 2D metal-insulator-transition 
(M-I-T) phenomenon, has attracted a great deal of experimental and 
theoretical attention. Among the puzzling and interesting aspects 
of the 2D M-I-T observations is the  strong  
parallel (to the 2D layer) magnetic field (B) dependence of the measured 
low temperature (T) resistivity $\rho$ of the apparent metallic phase at
electron (we will refer to both electrons and holes by the generic 
name ``electron'' in this paper) densities ($n$) above
the nominal M-I-T transition density ($n_c$). The unexpectedly strong 
parallel field (B in this paper refers exclusively to 
a 2D in-plane magnetic field with zero perpendicular field) dependence 
of the 2D resistivity  is suppressed deep into the 
metallic phase ($n\gg n_c$), and 
the system behaves more like a ``conventional metal''. 
The effect of B is dramatic \cite{one,two,three,four,five,six}: 
Modest B fields ($5-10T$) spectacularly increase the measured resistance 
by upto two orders of magnitude at low $n$, 
independent of whether the system is in the nominal metallic 
($n\ge n_c$) or insulating ($n \le n_c$) phase --- in Si inversion layer
$\rho(B)$ eventually seems to ``saturate'' at very high resistances
of the order of $\rho \sim  10^6 \Omega$ (in the nominal ``metallic'' phase)
for $B \sim 10 T$ whereas 
in p-GaAs systems no such resistivity saturation has yet been reported. 
In addition to this giant magnetoresistance 
the temperature dependence of $\rho(T)$ is 
also affected by B --- in particular, increasing B modifies the sign 
of $d\rho/dT$ which is positive (``metallic'' behavior) at low B and 
negative (``insulating'' behavior) at high B.

In this Letter we develop a possible theoretical explanation for the observed 
parallel magnetic field dependence of the 2D resistivity at low T and 
$n$. Our theoretical explanation for the observed
\cite{one,two,three,four,five,six} giant 
magnetoresistance and the associated ``suppression'' of the 
metallic phase is compellingly simple conceptually, and is robust in
the sense that 
it is independent of whether the conducting 
state is a true $T=0$ novel 2D metallic phase or just an effective ``metal'' 
(at finite T) by virtue of the localization length being exceedingly large. 
Our calculated B-dependent resistivity  is in 
good qualitative  agreement with the experimental 
observations \cite{one,two,three,four,five,six}, particularly in the
GaAs-based 2D systems, and can be systematically 
improved in the future.
The agreement between our theory and experimental 
results in Si inversion layers is not particularly good, and we mostly 
concentrate on GaAs systems in this paper.
In this Letter we use a minimal model  implementing the 
basic theoretical idea to obtain the generic trends in $\rho(B)$ in order 
to demonstrate  that we have found a possible explanation,
at least in the GaAs hole systems \cite{three,six}, for 
the giant magnetoresistance reported in recent 
experiments \cite{one,two,three,four,five,six}. We also emphasize that 
the theory presented in this Letter is totally independent of our $B=0$ 
theory for $\rho(T,n)$ developed in Ref. \onlinecite{seven}, and the 
parallel field results presented herein are {\it not} an extension and/or 
application of our $B=0$ theory.

The new physics 
we introduce in this paper is {\it the coupling 
of the parallel B-field to the orbital motion of the 2D carriers}. 
It has so far been assumed \cite{nine} that the observed parallel field 
induced giant magnetoresistance must {\it necessarily} be a spin effect 
because the orbital motion does not couple to a B field 
parallel to the 2D plane. This would certainly be true if the systems 
being studied were strictly 2D systems with zero thickness (i.e., perfect 
$\delta$-function confinement) in the direction (z) normal to 2D (x-y) 
plane. In reality, however, the systems being studied are 
{\it quasi}-2D  with their average thickness in the 
z-direction $\langle z \rangle$, being of the order of $30-300\AA$ 
depending on the 
system, carrier density, and other parameters (e.g., depletion charge
density) which are often not accurately known. Therefore, a parallel 
magnetic field in the x-y plane does couple to the z-orbital degree of 
freedom of the system, and such an orbital coupling could in fact be 
strong when $l_c < \langle z \rangle$ where $l_c = (\hbar c/eB)^{1/2}$ 
is the magnetic 
length  associated with the parallel field. Since $\langle z \rangle$ 
depends on (increases with decreasing $n$) the 
electron density $n$ \cite{eleven,twelve} through the self-consistent 
confinement potential, and may be quite large at low electron densities, 
the condition $l_c < \langle z \rangle$ is fulfilled in the low density 
experimental 
regime where the phenomenon of giant magnetoresistance is 
observed. In addition, the effect of the parallel field is
enhanced at low carrier densities
by the fact that the 2D Fermi wavelength $\lambda_F$ ($=2\pi/k_F$) 
at $B=0$ is substantially larger than the magnetic length $l_c$ associated 
with the typical B indicating a massive nonperturbative orbital 
effect associated with the applied parallel field. For example, for 
$B \approx 5T$ and $n=5\times 10^{10} cm^{-2}$, $\lambda_F \approx 1600 
\AA$ (Si); 1100 $\AA$ (p-GaAs), and $l_c \approx 100 \AA$ --- thus 
$\lambda_F \approx 10-15 \; l_c$, a situation which 
would be  achieved in a simple 
metal only for astronomical fields around $B\approx 10^7 T$. 
One should therefore expect a strong orbital effect 
arising from the applied field as reflected in 
the observed giant magnetoresistance in these systems. 
Since we are 
interested only in the orbital motion we neglect any Zeeman spin 
splitting in our consideration assuming a spin degeneracy of 2 throughout.

We note that the new physics arising from the coupling of the orbital 
motion to the parallel field in these quasi-2D inversion layer type 
systems has no analog in purely 3D or 2D systems. The zero 
field z-motion in these quasi-2D systems is quantized into subbands 
\cite{eleven,twelve} due to the confining potential --- the coupling 
of the B field to the orbital motion thus necessarily involves 
intersubband dynamics or scattering, and this intersubband dynamics 
gets coupled with the in-plane 2D dynamics in the presence of the 
parallel field, leading to the giant magnetoresistance. The effect 
should in fact persist to high densities (i.e., deep in the so-called 
metallic phase) except the size of the 
magnetoresistance should be only a few percent at high electron 
densities ($l_c \gg \langle z \rangle, \lambda_F$) whereas in the 
nonperturbative strong field limit the effect can be huge. Another 
way of understanding this strong (and unusual) orbital coupling is 
to note that the cyclotron energy associated with $B \approx 10 T$ 
is about 6 meV, which is larger than the zero field subband splitting 
and much larger than the 2D Fermi energy of the system in the low 
density regime. This interplay 
of cyclotron and subband dynamics is a novel feature of the quasi-2D 
system which has no purely 2D or 3D analog.
The physics of orbital coupling introduced in this paper could be 
thought of as a parallel field induced effective 2D to 3D crossover 
in the low density quasi-2D systems.

We implement the above idea by assuming a simple harmonic confinement 
of the z wavefunction at $B=0$. We adjust the parabolic
confinement potential variationally to obtain the best fit to the $B=0$ 
wavefunction and energy of the actual system at the appropriate density $n$. 
The harmonic confinement allows us to incorporate the effect of the 
parallel B field nonperturbatively \cite{twelve}, and in the high field 
limit ($l_c \ll \langle z \rangle_{B=0}$) our harmonic approximation 
(the Fock-Darwin levels) becomes almost 
exact. Any error arising from our somewhat inaccurate choice of the 
z-wavefunction at $B=0$ has only a small effect on the rather large 
magnetoresistance  we calculate. We take the B-field 
direction to be the x-axis without any loss of generality, and denote 
by $\rho_{xx}$ ($\rho_{yy}$) the 2D resistivity associated with the 
electric current flowing along (perpendicular) to the direction of B. 
As we describe below in details one of our specific predictions is that 
$\rho_{yy} \gg \rho_{xx}$ due to the large 2D anisotropy 
introduced by the applied B-field, provided there is no 
other anisotropy in the system.

For our transport calculation we apply the Boltzmann theory (in 
presence of the parallel field which is treated 
nonperturbatively by including it explicitly in our one electron 
wavefunction and energy via the harmonic confinement model \cite{twelve}) 
assuming scattering by short-range random impurities distributed uniformly 
throughout the quasi-2D layer.  Our neglect of screening effects 
\cite{seven} and the associated assumption of $\delta$-function impurity 
scattering potential arising from random impurity centers is done purely 
for the purpose of simplicity and  in order to keep the number of
free parameters at a minimum.
Since the effect we are considering is a rather gross 
effect (involving a very large increase in magnetoresistance) any errors 
associated with our simplified  scattering model are not of any 
qualitative significance in understanding the basic phenomenon.
One will have to improve the model (both for confinement and for impurity 
scattering) if one is interested in precise quantitative agreement with the 
experimental data in a specific system --- we believe  that such 
improvements
may be computationally extremely demanding.
Details of our calculation will be given in a future long publication.

We show our calculated results in Figs. 1 and 2 concentrating on the 
p-GaAs samples of Ref. \onlinecite{six}. In Fig. 1(a) we show our 
calculated $\rho_{xx}$ at $T=50 mK$ for various carrier densities as a 
function of the applied B-field. The harmonic confinement at each density 
has been variationally adjusted to give the best wavefunction for the 
holes in the GaAs heterostructure appropriate for that density at $B=0$.
The overall qualitative trends of our calculated results agree  well with 
the experimental data (c.f., Fig. 3 of Ref. \onlinecite{six}).
In fact, the specific experimental results of Ref. \onlinecite{six} 
agree very well with our calculations as can be seen in our Fig. 1(a) 
where we have put some representative experimental data.
At low B ($\omega_c < \omega_0$), $\ln (\rho_{xx})$ shows a $B^2$ dependence,
changing to a linear B dependence  
at high B ($\omega_c > \omega_0$) in agreement with experiment \cite{six}, 
where $\omega_c = eB/mc$ is the cyclotron frequency of the B field and 
$\omega_0$ is the curvature (or the subband splitting)
of the confinement potential. We note 
that the overall resistivity scale in our results is set by the density
$N_b$ of the $\delta$-function impurity scatterers which uniquely defines 
the $B=0$ value of $\rho_{xx}$ in a particular sample.
In Fig. 1(b) we show our calculated 
qualitative behavior for the Si inversion layer situation where the 
impurity scattering centers, instead of being randomly distributed 
throughout the layer, are located at the $Si-SiO_2$ inter-

\begin{figure}
\epsfysize=5.in
\epsffile{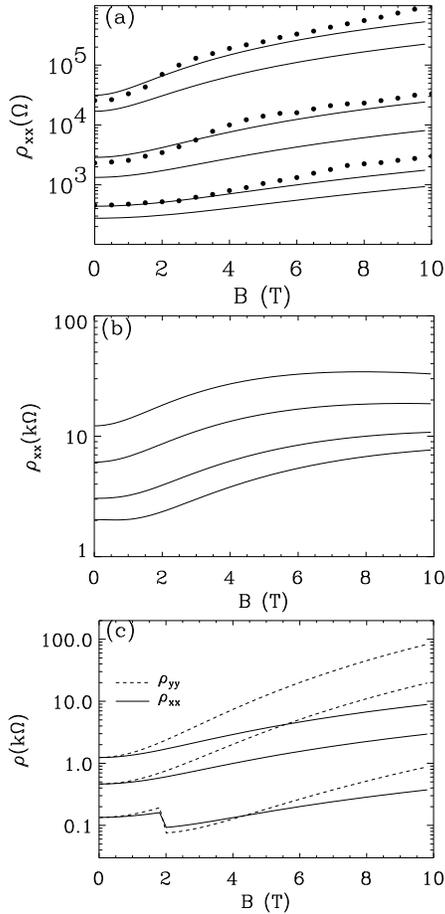}
\caption{
(a) Shows the calculated 2D resistivity $\rho_{xx}$ along the 
field direction as a function of the parallel field B at $T=50 mK$ 
in GaAs hole systems (appropriate for the samples in Ref. [6])
for $n=$ 0.89, 1.1, 1.63, 2.12, 3.23, 
4.11$\times 10^{10} cm^{-2}$ (top to bottom) with the confinement 
curvature $\omega_0=$ 0.85, 1.0, 1.35, 1.7, 2.3, 2.6 meV 
(top to bottom). Experimental points from Ref. [6] are shown as
dots for comparison. Random $\delta$-impurities are distributed uniformly 
through the layer. 
(b) Calculated $\rho_{xx}-B$ plot when the random $\delta$-impurities
are at the interface plane 85$\AA$ from the center of the Gaussian 
ground state wavefunction (at $B=0$). The top to bottom curves are 
for $n=$ 1, 2, 4 ,6$\times 10^{10}cm^{-2}$ and $\omega_0=1meV$. The 
saturation shifts to higher B values for higher densities.
(c) Calculated $\rho_{xx}-$ and $\rho_{yy}-B$ plots showing 
the strong 2D anisotropy with $\omega_0=1.5meV$ and $n=$1.5, 
4.0, 10.0$\times 10^{10}cm^{-2}$.
}
\end{figure}

\noindent
face or mostly on the insulating side.
In this situation (when the dominant scattering mechanism is 
planar, as due to the charged impurities at the interface and the 
surface roughness scattering at the $Si-SiO_2$ 
interface) elementary considerations show that the scattering effect 
must saturate eventually at high enough B fields as can be seen in 
Fig. 1(b) with the approximate saturation field increasing with increasing 
electron density consistent with experimental observations 
\cite{one,two,four,five}. The qualitative
difference between the non-saturation 
behavior [Fig. 1(a)] and the saturation behavior [Fig. 1(b)] arises from 
the $\delta$-function random scattering centers being distributed three 
dimensionally in Fig. 1(a) and in a 2D plane at the interface in Fig. 1(b). 
The results in Fig. 1(a) correspond qualitatively to GaAs where the main 
resistive scattering centers are the random impurities in GaAs whereas 
the results in Fig. 1(b) correspond more to Si inversion layer where 
the main scattering centers (charged impurities and surface roughness) 
are located in a plane near the interface.
We have not adjusted the parameters used for Fig. 1(b) to get agreement 
with Si inversion layer data --- our only point here is to demonstrate 
the qualitative physics underlying the saturation behavior.
In principle, we can get semi-quantitative agreement with any given 
set of data by adjusting the confinement parameter $\omega_0$, but 
we do not believe such an exercise to be meaningful particularly 
for the highly simplified model used in our theoretical calculations.

In Fig. 1(c) we show our predicted anisotropic magnetoresistivity with 
$\rho_{yy} \gg \rho_{xx}$ --- note that 
the anisotropy can be very large,  and decreases 
with increasing density. 
(The corresponding 2D Fermi surfaces, not shown here due to lack of 
space, are strongly anisotropic in shape, being elliptic rather than 
circular with the eccentricity increasing with increasing field.)
The highest density results (the lowest set of curves) in Fig. 1(c) show 
another predicted feature of our theory: At relatively high density, if 
the first excited subband of the system is occupied by the carriers at 
$B=0$ (a situation which in principle is achievable), 
the calculated resistivity would in fact first exhibit a negative 
magnetoresistance, as the excited subband depopulates with increasing B, 
before showing the characteristic giant magnetoresistance phenomenon. 
This oscillatory feature in the lowest set of curves in fig. 1(c) is 
the analog of the usual SdH oscillations in this problem. It 
is important to mention that the features predicted in Fig. 1(c) have 
already been {\it observed experimentally} \cite{thirteen} in parabolic 
{\it n-type} GaAs structures at higher densities (where the overall 
magnetoresistivity is a factor of 6 for $\rho_{yy}$ and only a factor 
of 2 for $\rho_{xx}$), and our calculations for this structure 
\cite{thirteen} agree well with the experimental findings.

It is important to emphasize that our predicted anisotropy (Fig. 1(c)),
which has been observed in n-type GaAs systems \cite{thirteen}, is {\it
not} seen \cite{one,fourteen} in Si inversion layers where $\rho_{xx} \sim
\rho_{yy}$ even in the presence of a strong applied parallel magnetic 
field. In this context we also point out that the saturation behavior 
of Fig. 1(b) calculated in our theory is not in particularly good 
agreement with the observed behavior \cite{one,two,four,five} in Si 
inversion layers where the saturation sets in more abruptly than in 
our theory. On the other hand our calculated magnetoresistance for 
GaAs systems, as shown in Fig. 1(a), is in excellent qualitative 
agreement with the corresponding GaAs results reported in Ref. 
\onlinecite{six}. The reason for this difference between the observed 
experimental behavior

\begin{figure}
\epsfysize=5.cm
\epsffile{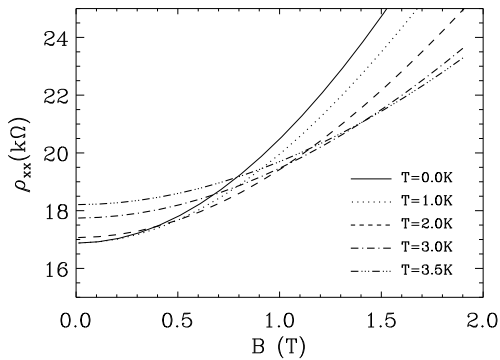}
\caption{
Calculated $\rho_{xx}-B$ plots for various T values as shown with 
$n=1.1\times10^{10}cm^{-2}$, $\omega_0 = 1.0 meV$.
}
\end{figure}

\noindent
between Si and GaAs based systems is currently 
not known. Our current theory applies rather well to the GaAs-based 
2D 
systems but not to Si systems for reasons which are not clear at 
this stage. One possibility for the difference between the two 
systems could be spin effects neglected in our theory.

Finally in Fig. 2 we provide the calculated temperature dependence of 
$\rho(T,B,n)$ which shows interesting non-monotonicity (``the suppression 
of the metallic phase'') experimentally \cite{six}. We emphasize that 
since we neglect all screening effects (due to our $\delta$-function 
impurity scattering model) our calculated temperature dependence, which 
arises entirely from Fermi surface and thermal occupancy effects, is 
necessarily simplistic (compared, for example with our $B=0$ theory 
for $\rho(T,n)$ as given in Ref. \onlinecite{seven}). Nevertheless we 
believe that even this drastically simplified model catches the basic 
physics of the phenomenon, and explains on a qualitative level why 
$d\rho/dT$ changes sign from being positive (``metallic'') at low field 
to negative (``insulating'') at high fields.
This is essentially a ``quantum-classical crossover'' type 
phenomenon \cite{seven} where the strong modification of the Fermi 
surface, as shown in Fig. 1(d), leads to non-monotonic temperature 
dependence at various B values. The physics of the negative $d\rho/dT$ 
at high fields  is entirely a Fermi surface effect. In 
Fig. 2 we plot $\rho_{xx}(B)$ for various fixed values of T, and 
we note that we obtain rough qualitative agreement with the experimental 
observations \cite{three,six} that there is a transition [around 
$B_c \sim 1T$ in Fig. 2] point $B_c$ where $d\rho(B)/dT$ changes 
its sign from being positive (``metallic'') for $B < B_c$ to being 
negative (``insulating'') for $B >B_c$ --- in our theoretical results 
$B_c$ is {\it not} a sharp transition point, rather a rough transition 
regime whereas in the experiment $B_c$ seems to be a sharp point. We 
see no particular reason for $B_c$ to be a sharp single transition point 
since this phenomenon is obviously not a phase transition (both $B < B_c$ 
and $B > B_c$ are effective ``metallic'' phases), and the crossover
behavior is only a Fermi surface (which is drastically distorted by 
high B values) effect. We suggest more precise measurements to check 
whether $B_c$ is really a single transition point or more a rough 
transition regime. 

We conclude by summarizing our theory and by briefly discussing our 
various approximations and limitations. We have shown that the observed 
giant positive 
magnetoresistance phenomenon in quasi-2D systems in the presence of a 
parallel magnetic field can be qualitatively
explained as arising from the coupling of the field 
to the carrier orbital motion by virtue of the finite thickness and the 
low density of the layer (spin plays no role in our explanation).
We predict a large anisotropy 
of resistivity in the 2D plane. Our
main approximations are: (1) Boltzmann transport theory; (2) harmonic 
confinement; (3) $\delta$-function random impurity scattering.
None of these approximations is qualitatively significant 
because we predict a very large (factors of $10-1000$) and robust effect. 
One important corollary of our theory is that the same effect should 
persist on the insulating side as long as the localization length is 
larger than the magnetic length except for the fact that $d\rho/dT$ 
should always be negative on the insulating side, which is
what is experimentally observed. Two important approximations
of our theory are that we have neglected all spin-related effects
as well as all crystallographic anisotropy effects, which could, in
principle, be added to our theory if future experiments demand such 
an improvement.
The main (and an important) limitation of the theory is that our 
predicted anisotropy seems {\it not} to be consistent with the 
existing data in Si inversion layers \cite{fourteen} where no 
magnetoresistive anisotropy has been seen. 
We do not know the reason for this disagreement --- one 
possibility being that there is an additional scattering mechanism, 
possibly spin-related, which also plays a role in the observed 
magnetoresistance and compensates for the anisotropy arising from 
orbital effects. The other possibility is that screening could be 
anisotropic in the presence of a strong 
parallel field (since the 2D Fermi surface is highly anisotropic), 
leading to a cancellation of the transport anisotropy in the Si 
inversion layer where screening effects are typically 
very strong (our theory neglects screening as we 
assume short range impurity scattering).
While more work is clearly needed to understand the quantitative details
of the observed magnetoresistance (particularly in Si inversion layers)
and spin-related effects may very well be playing an additional
role in the experimental data, our work compellingly demonstrates
the importance of {\it orbital} magnetoresistance in the presence
of a parallel magnetic field in the low density limit which
cannot be neglected in future work on the subject.

This work is supported by the U.S.-ARO and the U.S.-ONR.


\begin{thebibliography}{99}
\bibitem{nine} We note that each experimental publication
in refs. \onlinecite{one,two,three,four,five} above has an explicit 
statement categorically asserting that the observed phenomenon must 
be a spin (and not an orbital) effect, e.g., ``we note that a parallel 
magnetic field couples only to the spins of the electrons and not to 
their orbital motion.'' (quoted from ref. \onlinecite{five}).
\end{thebibliography}
\end{document}